\documentclass{article}
\usepackage{spconf,amsmath,graphicx}

\usepackage[nolist,nohyperlinks]{acronym}
\usepackage{multirow}
\usepackage[T1]{fontenc}
\usepackage{xcolor}
\usepackage{url}

\acrodef{DFL}{Deep Feature Loss}
\acrodef{SV}{Speaker Verification}
\acrodef{SSITW}{Simulated Speakers In The Wild}
\acrodef{SOTA}{State Of The Art}
\acrodef{CAN}{Context Aggregation Network}
\acrodef{SRE}{Speaker Recognition Evaluation}
\acrodef{VoxSRC}{VoxCeleb Speaker Recognition Challenge}
\acrodef{SEGAN}{Speech Enhancement Generative Adversarial Network}
\acrodef{GAN}{Generative Adversarial Network}
\acrodef{MMSE-GAN}{Minimum-Mean Squared Generative Adversarial Network}
\acrodef{PESQ}{Perceptual Evaluation of Speech Quality}
\acrodef{SDR}{Signal-to-Distortion Ratio}
\acrodef{LSTM}{Long-Short Term Memory}
\acrodef{ASR}{Automatic Speech Recognition}
\acrodef{WER}{Word Error Rate}
\acrodef{SOTA}{state-of-the-art}
\acrodef{PLDA}{Probabilistic Linear Discriminant Analysis}
\acrodef{TSE}{Temporal Squeeze Excitation}
\acrodef{TF}{Time-Frequency}
\acrodef{BN}{Batch Normalization}
\acrodef{EDN}{Encoder-Decoder Network}
\acrodef{IN}{Instance Normalization}
\acrodef{RAdam}{Rectified Adam}
\acrodef{MFCC}{Mel-Frequency Cepstrum Coefficient}
\acrodef{LDE}{Learnable Dictionary Encoding}
\acrodef{ETDNN}{Extended TDNN}
\acrodef{FTDNN}{Factorized TDNN}
\acrodef{SNR}{Signal-to-Noise Ratio}
\acrodef{WADASNR}{Waveform Amplitude Distribution Analysis}
\acrodef{SITW}{Speakers In The Wild}
\acrodef{MVN}{Mean-Variance Normalized}
\acrodef{EER}{Equal Error Rate}
\acrodef{minDCF}{Minimum Decision Cost Function}
\acrodef{AHC}{Agglomerative Hierarchical Clustering}
\acrodef{TDNN}{Time-Delay Neural Network}

\newcommand{\comment}[1]{}

\title{Feature Enhancement with Deep Feature Losses\\for Speaker Verification}
\name{Saurabh Kataria,
Phani Sankar Nidadavolu,
Jes\'us Villalba,
Nanxin Chen,
Paola Garc\'ia-Perera,
Najim Dehak
\thanks{The research reported here was conducted at the 2019 Frederick Jelinek Memorial Summer Workshop on Speech and Language Technologies, hosted at L'\'Ecole de Technologie Sup\'erieure (Montreal, Canada) and sponsored by Johns Hopkins University with unrestricted gifts from Amazon, Facebook, Google, and Microsoft.}}
\address{Center for Language and Speech Processing, Johns Hopkins University, Baltimore, MD, USA}
\begin{document}
%
\maketitle

\begin{abstract}
Speaker Verification still suffers from the challenge of generalization to novel adverse environments.
We leverage on the recent advancements made by deep learning based speech enhancement and propose a feature-domain supervised denoising based solution.
We propose to use Deep Feature Loss which optimizes the enhancement network in the hidden activation space of a pre-trained auxiliary speaker embedding network.
We experimentally verify the approach on simulated and real data.
A simulated testing setup is created using various noise types at different SNR levels.
For evaluation on real data, we choose BabyTrain corpus which consists of children recordings in uncontrolled environments.
We observe consistent gains in every condition over the state-of-the-art augmented Factorized-TDNN x-vector system.
On BabyTrain corpus, we observe relative gains of 10.38\% and 12.40\% in minDCF and EER respectively.
\end{abstract}


\vspace{-1.5em}
\section{Introduction}
\label{sec:intro}
Various phenomena degrades speech such as noise, reverberation, speaker movement, device orientation, and room characteristics~\cite{kataria2017hearing}.
This makes the deployment of \ac{SV} systems challenging.
To address this, several challenges were organized recently such as NIST \ac{SRE} 2019, VOiCES from a Distance Challenge~\cite{nandwana2019voices}, and \ac{VoxSRC} 2019.
We consider acoustic feature enhancement as a solution to this problem.
In the past decade, deep learning based enhancement has made great progress.
Notable approaches include mask estimation, feature mapping, \ac{GAN}~\cite{pascual2017segan}, and \ac{DFL}~\cite{germain2018speech}.
Usually, such works report on enhancement metrics like \ac{PESQ} and \ac{SDR} on small datasets like VCTK.
Some works tackle joint denoising-dereverberation, unsupervised enhancement, and source separation.
However, we focus on supervised denoising. Specifically, we are interested in enhancement for improving the robustness of other speech \emph{tasks}.
We refer to this methodology as \emph{task-specific} enhancement.

\emph{Task-specific} enhancement has been proposed for \ac{ASR}, Language Recognition, and \ac{SV}.
We focus on single-channel wide-band \ac{SV}, for which augmented x-vector network with \ac{PLDA} back-end is the \ac{SOTA}~\cite{villalba2019state}.
For \ac{SV}, \cite{shon2019voiceid} and \cite{michelsanti2017conditional} have reported improvements on simulated data.
We note that x-vector systems still face significant challenge in adverse scenarios, as demonstrated in a recent children speech diarization study~\cite{xie2019multi}.
This interests us in investigating if \emph{task-specific} enhancement can complement \ac{SOTA} x-vector based \ac{SV} systems.

We argue that the training of \emph{task-specific} enhancement system should depend on the \emph{task}.
Therefore, we build on the ideas of Perceptual Loss \cite{johnson2016perceptual} and propose a solution based on the speech denoising work in \cite{germain2018speech}.
In \cite{germain2018speech}, authors train a speech denoising network by deriving loss from a pre-trained speech classification network.
There are several differences in our work from \cite{germain2018speech}.
First, we choose the auxiliary task same as the x-vector network task i.e. speaker classification.
This follows from the motivation to use \emph{task-specific} enhancement to improve upon the \ac{SOTA} x-vector system for \ac{SV}.
Second, we enhance in feature-domain (log Mel-filterbank), which makes it conducive for use with \ac{MFCC} based auxiliary network.
Lastly, we demonstrate the proof-of-concept using datasets of much larger scale.
An added advantage of our proposed approach is that we do enhancement only during inference, thus, avoiding the need for re-training of x-vector network.

\section{Deep Feature Loss}
\label{sec:dfl}
Perceptual Loss or \emph{deep feature loss} refers to use of a pre-trained auxiliary network for the training loss.
The auxiliary network is trained for a different task and returns loss in form of hidden layer activations from multiple layers.
In~\cite{germain2018speech}, authors train an enhancement system with an audio classification auxiliary network.
The loss is the $L_1$ deviation of the activations of clean and enhanced signal.
We refer to this as \emph{deep feature loss} (DFL), while \emph{feature loss} (FL) refers to the independent na\"ive training of enhancement system without auxiliary network.
For batch size of 1, the loss functions for DFL, FL, and DFL+FL (combination) are given below.
\begin{equation}
\begin{split}
    \mathcal{L}_{\text{DFL}}(F_n,F_c) &= \sum_{i=1}^L\mathcal{L}_{\text{DFL},i}(F_n,F_c) \\
        &= \sum_{i=1}^L||a_i(F_c) - a_i(e(F_n))||_{1,1}
\end{split}
\end{equation}
\begin{equation}
    \mathcal{L}_{\text{FL}}(F_n,F_c) = ||F_c - e(F_n)||_{1,1}
\end{equation}
\begin{equation}
    \mathcal{L}_{\text{DFL+FL}}(F_n,F_c) = \mathcal{L}_{\text{DFL}}(F_n,F_c) + \mathcal{L}_{\text{FL}}(F_n,F_c)
\end{equation}

Here, $F_n$ and $F_c$ are $F \times T$ matrices containing features for the current pair of noisy and clean sample respectively. $F$ is the number of frequency bins, $T$ is the number of frames, $e(\cdot)$ is the enhancement network, $a(\cdot)$ is the auxiliary network, $a_i(\cdot)$ is the output of the $i$-th layer of $a(\cdot)$ considered for computing \ac{DFL}, and $L$ is the number of layers of $a(\cdot)$ whose outputs are used for computing \ac{DFL}. We fix the coefficients of $\mathcal{L}_{\text{DFL},i}(\cdot,\cdot)$ and $\mathcal{L}_{\text{FL}}(\cdot,\cdot)$ equal to 1. We tried the coefficient re-weighting scheme of \cite{germain2018speech} but found it unhelpful. $L$ depends on the architecture of $a(\cdot)$. We fix it to 6, as suggested by our preliminary experiments.

\section{Neural Network Architectures}

\subsection{Enhancement Networks}
Here, we describe the two \emph{fully-convolutional} architectures we designed as candidates for the enhancement network.

\vspace{-0.5em}
\subsubsection{Context Aggregation Network}
A deep CNN with dilated convolutions increases the receptive field of network monotonically, resulting in large temporal context.
In~\cite{germain2018speech}, authors design such a network for time-domain signal using 1-D convolutions.
The first layer of our \ac{CAN} is a 2-D \ac{BN} layer. It has eight 2-D convolution layers with kernel size of 3x3, channel dimension of 45, and dilation linearly increasing from 1 to 8.
Between CNN layers, is an Adaptive \ac{BN} layer followed by a LeakyReLU activation of slope 0.2. We introduced several modifications to the architecture in \cite{germain2018speech}. First, we include, uniformly separated, three \ac{TSE} connections along with residual connections.
\ac{TSE} is a variant of Squeeze Excitation~\cite{hu2018squeeze}, where instead of obtaining a global representation common to all \ac{TF} bins (by average pooling in both dimensions), we obtain a representation per frequency bin (pooling just in time dimension). Then, we 
 compute excitation weights for every \ac{TF} bin. Finally, a linear layer is used to map to original input dimension. The network output is assumed to represent a mask that we have multiply by the noisy features to obtain the clean features in linear domain. Since, we used acoustic features in $\log$ domain. We apply $\log$ the network output and add to the input to obtain the enhanced features in $\log$ domain.  The network has a context length of 73 frames and number of parameters are 2.6M.

\vspace{-0.5em}
\subsubsection{Encoder-Decoder Network}
We modify the \ac{EDN} architecture of the generator of Cycle-GAN in the domain adaptation work of~\cite{nidadavolu2019lr,nidadavolu2019unsupervised}.
\ac{EDN} has several residual blocks after the encoder and a skip connection.
Details can be found in \cite{nidadavolu2019cycle}.
We make three modifications.
First, the number of channels are set to a high value of 90.
Second, Swish activation function \cite{ramachandran2017swish} is used instead of ReLU.
Lastly, the training details are different, particularly, in the context of optimization (refer Section \ref{subsec:train}). The network has a context length of 55 and number of parameters are 22.5M.

\vspace{-0.5em}
\subsection{Speaker Embedding Networks}
\subsubsection{Residual Network}
\label{subsubsec:res}
The auxiliary network in our \ac{DFL} formulation is the ResNet-34-LDE network described in~\cite{villalba2018jhu,snyder2019jhu,villalba2019state}. 
It is a ResNet-34 residual network with \ac{LDE} pooling and Angular Softmax loss function. The dictionary size of \ac{LDE} is 64 and the network has 5.9M parameters.

\vspace{-0.5em}
\subsubsection{x-vector Network}
We experiment with two x-vector networks, \ac{ETDNN} and \ac{FTDNN}. \ac{ETDNN} improves upon the previously proposed \ac{TDNN} system by interleaving dense layers in between the convolution layers. The \ac{FTDNN} network forces the weight matrix between convolution layers to be a product of two low rank matrices. Total parameters for \ac{ETDNN} and \ac{FTDNN} are 10M and 17M respectively. A summary of those networks can be found in~\cite{villalba2019state}.

\section{Experimental Setup}

\subsection{Dataset Description}
We combine VoxCeleb1 and VoxCeleb2 \cite{chung2018voxceleb2} to create \textit{voxceleb}.
Then, we concatenate utterances extracted from the same video to create \textit{voxcelebcat}.
This results in 2710 hrs of audio with 7185 speakers.
A random 50\% subset of \textit{voxcelebcat} forms \textit{voxcelebcat\_div2}.
To ensure sampling of clean utterances (required for training enhancement), an SNR estimation algorithm (\ac{WADASNR} \cite{kim2008robust}) is used to sample top 50\% clean samples from \textit{voxcelebcat} to create \textit{voxcelebcat\_wadasnr}.
This results in 1665 hrs of audio with 7104 speakers.
To create the noisy counterpart, MUSAN \cite{snyder2015musan} and DEMAND \cite{thiemann2013diverse} are used.
A 90-10 split gives us a parallel copy of training and validation data for the enhancement system.
The auxiliary network is trained with \textit{voxcelebcat\_wadasnr}.
Lastly, \textit{voxcelebcat\_combined} is formed by data augmentation with MUSAN to create a dataset of size three times \textit{voxcelebcat}.

We design a simulated testing setup called \ac{SSITW}. Several noisy test sets are formed by corrupting \ac{SITW} \cite{mclaren2016speakers} core-core condition with MUSAN and ``background noises'' from CHiME-3 challenge (referred to as \textit{chime3bg}). This results in five test SNRs (-5dB, 0dB, 5dB, 10dB, 15dB) and four noise types (\textit{noise}, \textit{music}, \textit{babble}, \textit{chime3bg}). Here, \textit{noise} refers to ``noise category'' in MUSAN, consisting of common environmental acoustic events. It is ensured that the test noise files are disjoint from the training ones.

We choose \emph{BabyTrain} corpus for evaluation on real data. It is based on the Homebank repository \cite{vandam2016homebank} and consists of daylong children speech around other speakers in uncontrolled environments.
Training data for diarization and detection (\textit{adaptation data}) are around 130 and 120 hrs respectively, while enrollment and test data are around 95 and 30 hrs respectively. This data was split into enrollment and test utterances which
were classified as per their duration.
In our terminology, \textit{test>=$n$ sec} and \textit{enroll=$m$ sec} refers to test and enrollment utterances of minimum $n$ and equal to $m$ seconds from the speaker of interest respectively with
$n \in \{0,5,15,30\}$ and $m \in \{5,15,30\}$. For enrollment, time marks of the target speaker were given but not for test where multiple speakers may be present.

We now describe the training data for our three x-vector based baseline systems.
For the first (and simplest) baseline, we use \ac{ETDNN}. The training data for \ac{ETDNN} as well as its \ac{PLDA} back-end is \textit{voxcelebcat\_div2}. Since no data augmentation is done, we refer to this system as \emph{clean x-vector} system or \emph{ETDNN\_div2}. For the second and third baseline, we choose \ac{FTDNN}, which is trained with \textit{voxcelebcat\_combined} and several SRE datasets. Its details can be found in \cite{villalba2018jhu}. These two baselines are referred to as \emph{augmented x-vector} systems. The difference between the second (\emph{FTDNN\_div2}) and the third baseline (\emph{FTDNN\_comb}) is that they use \textit{voxcelebcat\_div2} and \textit{voxcelebcat\_combined} as \ac{PLDA} training data respectively. There is an additional \ac{PLDA} in the diarization step for \emph{BabyTrain}, for which \textit{voxceleb} is used.

\vspace{-0.5em}
\subsection{Training details}
\label{subsec:train}
\ac{CAN} is trained with batch size of 60, learning rate of 0.001 (exponentially decreasing), number of epochs as 6, optimizer as Adam, and 500 number of frames (5s audio).
The differences for \ac{EDN} is in batch size (32) and optimizer (\ac{RAdam}).
Differences arise due to the independent tuning of two networks.
However, they are both trained with unnormalized 40-D log Mel-filterbank features.
The auxiliary network is trained with batch size of 128, number of epochs as 50, optimizer as Adam, learning rate of 0.0075 (exponentially decreasing) with warmup, and sequences of 800 frames (8s audio).
It is trained with mean-normalized log Mel-filterbank features.
To account for this normalization mismatch, we do online mean normalization between the enhancement and auxiliary network.
\ac{ETDNN} and \ac{FTDNN} are trained with Kaldi scripts using \ac{MVN} 40-D \ac{MFCC} features.

\subsection{Evaluation details}
\label{subsec:evaldetails}

The \ac{PLDA}-based backend for \ac{SSITW} consists of a 200-D LDA with generative Gaussian SPLDA~\cite{villalba2018jhu}.
For evaluation on \textit{BabyTrain}, a diarization system is used additionally to account for the multiple speakers in test utterances.
We followed the Kaldi x-vector callhome diarization recipe.
Details are in the \textit{JHU-CLSP diarization system} described in~\cite{villalba2018jhu}.
Note that only test, enroll, and \textit{adaptation data} utterances were enhanced.
For the final evaluation, we use standard metrics like \ac{EER} and \ac{minDCF} at target prior $p=0.05$ (NIST SRE18 VAST operating point).
The Code for this work is available online \footnote{\url{https://github.com/jsalt2019-diadet}} and a parent paper is submitted in parallel \cite{garcia2019speaker}.

\section{Results}
\subsection{Baseline results}
In Table \ref{tab:baseline}, we present the baseline (averaged) results on simulation and real data.
As expected, \textit{clean x-vector} system performs worst. Among \ac{SSITW} and \textit{BabyTrain}, we observe different trends using the \textit{augmented x-vector} systems.
\textit{FTDNN\_div2} performs better for \textit{BabyTrain}, while \textit{FTDNN\_comb} performs better for \ac{SSITW}.
Due to focus on real data, we drop third baseline from further analysis.

\vspace{-3pt}
\begin{table}[htbp]
    \centering
    \caption{Baseline results using three verification systems}
    \resizebox{0.48\textwidth}{!}{
        \begin{tabular}[t]{|l|cc|cc|}
            \hline
            & \multicolumn{2}{c|}{SSITW} & \multicolumn{2}{c|}{BabyTrain} \\
            \cline{2-5}
            & EER & minDCF & EER & minDCF \\
            \hline
            ETDNN\_div2 & 10.75 & 0.608 & 13.90 & 0.783 \\
            FTDNN\_div2 & 5.70 & 0.357 & \textbf{7.66} & \textbf{0.366} \\
            FTDNN\_comb & \textbf{3.70} & \textbf{0.222} & 9.72 & 0.409 \\
            \hline
        \end{tabular}
    }
    \label{tab:baseline}
\end{table}
\vspace{-7pt}
\subsection{Comparison of Context Aggregation Network and Encoder-Decoder Network}

Table \ref{tab:CANvsEDN} present enhancement results using the two candidate enhancement networks.
There is a difference in performance trend among \ac{CAN} and \ac{EDN}.
On \ac{SSITW}, \ac{EDN} works better, while on \textit{BabyTrain}, \ac{CAN} gives better performance.
Again, due to focus on real data, \ac{CAN} is chosen for further analysis.
Results can be compared with Table \ref{tab:baseline} and the benefit of enhancement can be noted for both baseline systems.
Underlined numbers represent the overall best performance attained in this study for each dataset.

\vspace{-1.5em}
\begin{table}[htbp]
    \centering
    \caption{Comparison of enhancement by \ac{CAN} and \ac{EDN}}
    \resizebox{0.48\textwidth}{!}{
        \begin{tabular}{|c|c|cc|cc|}
            \hline
            & & \multicolumn{2}{c|}{SSITW} & \multicolumn{2}{c|}{BabyTrain} \\
            \cline{3-6}
            & & EER & minDCF & EER & minDCF \\
            \hline
            \multirow{3}{*}[1.25ex]{\rotatebox[origin=c]{90}{CAN}} & ETDNN\_div2 & 7.61 & 0.450 & 10.33 & 0.510 \\
                & FTDNN\_div2 & 5.37 & 0.333 & \textbf{6.71} & \textbf{0.328} \\
                \hline
            \multirow{3}{*}[1.25ex]{\rotatebox[origin=c]{90}{EDN}} & ETDNN\_div2 & 6.51 & 0.398 & 11.76 & 0.561 \\
                & FTDNN\_div2 & \underline{\textbf{4.18}} & \underline{\textbf{0.273}} & 7.35 & 0.334 \\
                \hline
        \end{tabular}
    }
    \label{tab:CANvsEDN}
\end{table}

\vspace{-1.75em}
\subsection{Comparison of feature and Deep Feature Loss}
Table \ref{tab:featvsDFL} present results using the three loss functions using the stronger baseline (\textit{FTDNN\_div2}). The loss function in our proposed solution ($\mathcal{L}_{\text{DFL}}$) gives best performance.
It is important to note that the na\"ive enhancement ($\mathcal{L}_{\text{FL}}$), which does not use auxiliary network, gives worse results than baseline.
Since we predict mask, $\mathcal{L}_{\text{FL}}$ is comparable with the mask-based enhancement in literature.
The combination loss ($\mathcal{L}_{\text{DFL+FL}}$) gives slightly better \ac{EER} on \textit{BabyTrain} but degrades all other metrics.
The last row represents the performance difference between the na\"ive and the proposed scheme.
In next sections, we present detailed results on both datasets using $\mathcal{L}_{\text{DFL}}$.
\begin{table}[htbp]
    \centering
    \caption{Comparison of three losses on \textit{FTDNN\_div2}}
    \begin{tabular}[t]{|l|cc|cc|}
        \hline
        & \multicolumn{2}{c|}{SSITW} & \multicolumn{2}{c|}{BabyTrain} \\
        \cline{2-5}
        & EER & minDCF & EER & minDCF \\
        \hline
        FTDNN\_div2 & 5.70 & 0.357 & 7.66 & 0.366 \\
        \hline
        $\mathcal{L}_{\text{FL}}$ & 8.51 & 0.516 & 7.90 & 0.485 \\
        $\mathcal{L}_{\text{DFL}}$ & \textbf{5.37} & \textbf{0.333} & \underline{\textbf{6.71}} & \underline{\textbf{0.328}} \\
        $\mathcal{L}_{\text{DFL+FL}}$ & 6.27 & 0.381 & 7.30 & 0.383 \\
        \hline
        $\mathcal{L}_{\text{FL}}-\mathcal{L}_{\text{DFL}}$ & 3.14 & 0.183 & 1.19 & 0.157 \\
        \hline
    \end{tabular}
    \label{tab:featvsDFL}
\end{table}

\subsection{Results on Simulated Speakers In The Wild}
In Table \ref{tab:SSITW}, we present results on \ac{SSITW} per noise condition. The upper half of table shows results with and without enhancement using \textit{clean x-vector}.
The performance gain in every condition is consistent.
We note here that the \textit{babble} condition is the most challenging.
The lower half of table shows results using the \textit{augmented x-vector}.
The performance gain is lesser albeit consistent here.
$\Delta$ (in \%) represents the relative change in metric after enhancement. Asterisk (*) denotes the metric value after enhancement.
\begin{table}[htbp]
    \centering
    \caption{Results with and without DFL enhancement on SSITW using two baseline systems}
    \resizebox{0.48\textwidth}{!}{
        \begin{tabular}{|l|c|cccc|}
            \hline
            & & noise & music & babble & chime3bg \\
            \hline
            \multirow{3}{*}[-1ex]{\rotatebox[origin=c]{90}{ETDNN\_div2}} & EER & 8.52 & 9.17 & 13.36 & 11.94 \\
                & EER* & 5.98 & 6.31 & 10.6 & 8.19 \\
                & $\Delta$ & -29.81\% & -31.19\% & -20.66\% & -31.41\% \\
                \cline{2-6}
                & minDCF & 0.546 & 0.552 & 0.661 & 0.672 \\
                & minDCF* & 0.381 & 0.391 & 0.544 & 0.484 \\
                & $\Delta$ & -30.22\% & -29.17\% & -17.70\% & -27.98\% \\
                \hline
            \multirow{3}{*}[-1.8ex]{\rotatebox[origin=c]{90}{FTDNN\_div2}} & EER & 3.80 & 4.42 & 8.75 & 6.49 \\
                & EER* & \textbf{3.69} & \textbf{3.83} & \textbf{8.06} & \textbf{5.88} \\
                & $\Delta$ & -2.90\% & -13.35\% & -7.89\% & -9.40\% \\
                \cline{2-6}
                & minDCF & 0.264 & 0.301 & 0.461 & 0.402 \\
                & minDCF* & \textbf{0.253} & \textbf{0.269} & \textbf{0.435} & \textbf{0.375} \\
                & $\Delta$ & -4.17\% & -10.63\% & -5.64\% & -6.72\% \\
                \hline
        \end{tabular}
    }
    \label{tab:SSITW}
\end{table}

\subsection{Results on BabyTrain}
In Table \ref{tab:BabyTrain}, we present results on \textit{BabyTrain} per test duration condition (averaged over all enroll durations).
Similar to the previous section, we observe high gains using the \textit{clean x-vector}.
The lower half of table also shows consistent significant improvement in every condition.
It is important to note that even with a strong \ac{FTDNN} based \textit{augmented x-vector} baseline, enhancement helps significantly.
Also, the easier the test condition, the higher the improvement.
\begin{table}[htbp]
    \centering
    \caption{Results with and without DFL enhancement on BabyTrain using two baseline systems}
    \resizebox{0.48\textwidth}{!}{
        \begin{tabular}{|l|c|cccc|}
            \hline
            & & test>=30s & test>=15s & test>=5s & test>=0s \\
            \hline
            \multirow{3}{*}[-1ex]{\rotatebox[origin=c]{90}{ETDNN\_div2}} & EER & 9.83 & 12.94 & 16.26 & 16.57 \\
                & EER* & 6.80 & 9.35 & 12.40 & 12.78 \\
                & $\Delta$ & -30.82\% & -27.74\% & -23.74\% & -22.87\% \\
                \cline{2-6}
                & minDCF & 0.673 & 0.782 & 0.837 & 0.840 \\
                & minDCF* & 0.378 & 0.517 & 0.581 & 0.587 \\
                & $\Delta$ & -43.83\% & -33.89\% & -30.59\% & -30.12\% \\
                \hline
            \multirow{3}{*}[-1.8ex]{\rotatebox[origin=c]{90}{FTDNN\_div2}} & EER & 4.67 & 6.50 & 9.54 & 9.92 \\
                & EER* & \textbf{3.97} & \textbf{5.67} & \textbf{8.41} & \textbf{8.78} \\
                & $\Delta$ & -14.99\% & -12.77\% & -11.84\% & -11.49\% \\
                \cline{2-6}
                & minDCF & 0.242 & 0.335 & 0.440 & 0.447 \\
                & minDCF* & \textbf{0.204} & \textbf{0.298} & \textbf{0.400} & \textbf{0.409} \\
                & $\Delta$ & -15.70\% & -11.04\% & -9.09\% & -8.50\% \\
                \hline
        \end{tabular}
    }
    \label{tab:BabyTrain}
\end{table}

\vspace{-10pt}
\section{Conclusion}
\label{sec:conclusion}
We propose to do feature-domain enhancement at the front-end of the x-vector based Speaker Verification system and claim that it improves robustness.
To establish the proof-of-concept, we experiment with two enhancement networks, three loss functions, three baselines, and two testing setups.
We create simulation data using noises of different types at a broad range of SNRs.
For evaluation on real data, we choose \textit{BabyTrain}, which consists of day-long children recordings in uncontrolled environments.
Using \textit{deep feature loss} based enhancement, we observe consistent gains in every condition of simulation and real data.
On \textit{BabyTrain}, we observe relative gain of 10.38\% in minDCF and 12.40\% in EER.
In future, we will explore our idea with more real noisy datasets.

\clearpage

\bibliographystyle{IEEEbib}
\bibliography{refs}

\end{document}